\documentclass[epj]{svjour}
\usepackage{epsf}
\usepackage{amsmath,amsbsy}
\usepackage{graphicx}

\numberwithin{equation}{section}
\begin{document}
\title{A Minimal Model for Vorticity and Gradient Banding in Complex Fluids}
\titlerunning{Vorticity and Gradient Banding in Complex Fluids}
\author{J.L.  Goveas \inst{1} and P.D. Olmsted \inst{2}}
\institute{Department of Chemical Engineering, MS 362 Rice University,
  6100 Main Street, Houston, TX 77005 \email{jlgoveas@rice.edu}\and
  Polymer IRC and 
  Department of Physics \& Astronomy, University of Leeds, Leeds LS2
  9LT, UK \email{p.d.olmsted@leeds.ac.uk}} \date{\today}
\abstract{A general phenomenological reaction-diffusion model for
  flow-induced phase transitions in complex fluids is presented.  The
  model consists of an equation of motion for a nonconserved
  composition variable, coupled to a Newtonian stress relations for
  the reactant and product species.  Multivalued reaction terms allow
  for different homogeneous phases to coexist with each other,
  resulting in banded composition and shear rate profiles.  The
  one-dimensional equation of motion is evolved from a random initial
  state to its final steady-state.  We find that the system chooses
  banded states over homogeneous states, depending on the shape of the
  stress constitutive curve and the magnitude of the diffusion
  coefficient. Banding in the flow gradient direction under shear rate
  control is observed for shear-thinning transitions, while banding in the
  vorticity direction under stress control is observed for shear-thickening
  transitions.
}
\maketitle
\section{Introduction}
There is a significant body of experimental evidence documenting the
existence of sharp, stable interfaces separating two or more phases or
``bands'', in shear flow in complex fluids.  This phenomena has been
reported in various types of surfactant solutions \cite{Pine,Bonn},
polymers \cite{Eiser}, liquid crystals \cite{neutron} 
and colloidal suspensions \cite{Zcolloids}.
There appears to be a compelling generality between these ``phase
transitions'' in different complex fluids:
\begin{itemize}
\item[i)] The onset of banding or phase separation manifests itself as
  a discontinuity in the ``flow curve'' of the system. The flow curve
  is the unique relationship between the measured shear stress and the
  applied shear rate (or vice versa) at steady-state.  An experimental
  flow curve typically contains segments that correspond to
  homogeneous flow, as well as segments corresponding to inhomogeneous
  flow.  The individual homogeneous bands which make up the
  inhomogeneous state each have their own homogeneous flow curve,
  which we shall refer to as a ``constitutive curve''.  The
  inhomogeneous flow curve then represents the response of the system,
    averaged over different spatial regions that occupy different
    homogeneous flow branches, in
    proportions to maintain the externally
  controlled shear stress or shear rate.
  [Henceforth we will use the terms stress and shear stress
  interchangeably, unless otherwise specified.]
\item[ii)] The transition only occurs above a unique and reproducible
  critical stress or shear rate.  
  \item[iii)] The flow curve can be qualitatively different depending
    on whether the average stress or the average shear rate in the
    system is held fixed.  [In a typical rheological
    experiment this is achieved by controlling the torque or angular
    velocity respectively.]  For 
    intermediate stresses or shear rates, the flow
    curve usually has multiple branches which are not equally accessible under
    both stress and shear rate control. For weak and strong
      flows, the flow curve  is
    single-valued, and the same locus of points is traced out under
    stress or shear rate control.
\item[iv)] The flow-induced bands have different shear rates or shear
  stresses, and are generally also distinguished by some
    combination of different degrees of order and different
    microstructures.
\item[v)]The interfaces between the bands may be aligned in the
  direction of the flow gradient or the flow vorticity. Each banding
  orientation has its own rheological signature. In shear-thinning
  systems, for example, a stress plateau in the flow curve usually
  indicates gradient banding, while extrema in the stress (as a
  function of shear rate) usually indicate vorticity banding.  One of
  us \cite{Peter} has constructed possible flow curves based on the
  banding orientation and the character of phase coexistence
  (shear-thinning versus shear-thickening).
\end{itemize}

Gradient banding has been unambiguously observed in solutions of
wormlike micelles. In strain-controlled experiments on shear-thinning
solutions, a stress plateau coincides with shear-banding in the
gradient direction \cite{Callaghan}.  In shear-thickening solutions
\cite{Pine}, a gel-like phase can be induced by flow.  Under shear
rate control the induced phase fills the system at steady-state,
resulting in a discontinuous stress jump in the flow curve.  Under
stress control phase coexistence between solution and gel is observed,
the gel fraction being an increasing function of stress. In
  the corresponding flow curve the shear rate
shows a minimum and maximum. 

Vorticity banding has been reported in dense colloidal suspensions
\cite{Zcolloids} and surfactant solutions of multilamellar vesicles
\cite{Bonn}.  When the shear rate is held fixed, the flow curve shows
a maximum and minimum in the stress. Under controlled stress, there is
a jump up in shear rate upon increasing stress, and a jump down in
shear rate upon decreasing stress.  The same qualitative curves have
also been observed in surfactant hexagonal phases \cite{LRamos},
although in that case vorticity banding has not yet been explicitly
verified.  Such behavior is analogous to that of the shear-thickening
wormlike micelles, if the roles of stress and shear rate are
interchanged.  In bcc cubic crystals of triblock copolymers
\cite{Eiser}, Eiser \textit{et al.} observe two stress plateaus in the
flow curve under controlled shear rate. X-ray diffraction shows that
each plateau corresponds to different orientations (relative to the
flow direction) of dense planes in the crystal.

\begin{figure}
\centering{\includegraphics[scale=0.6]{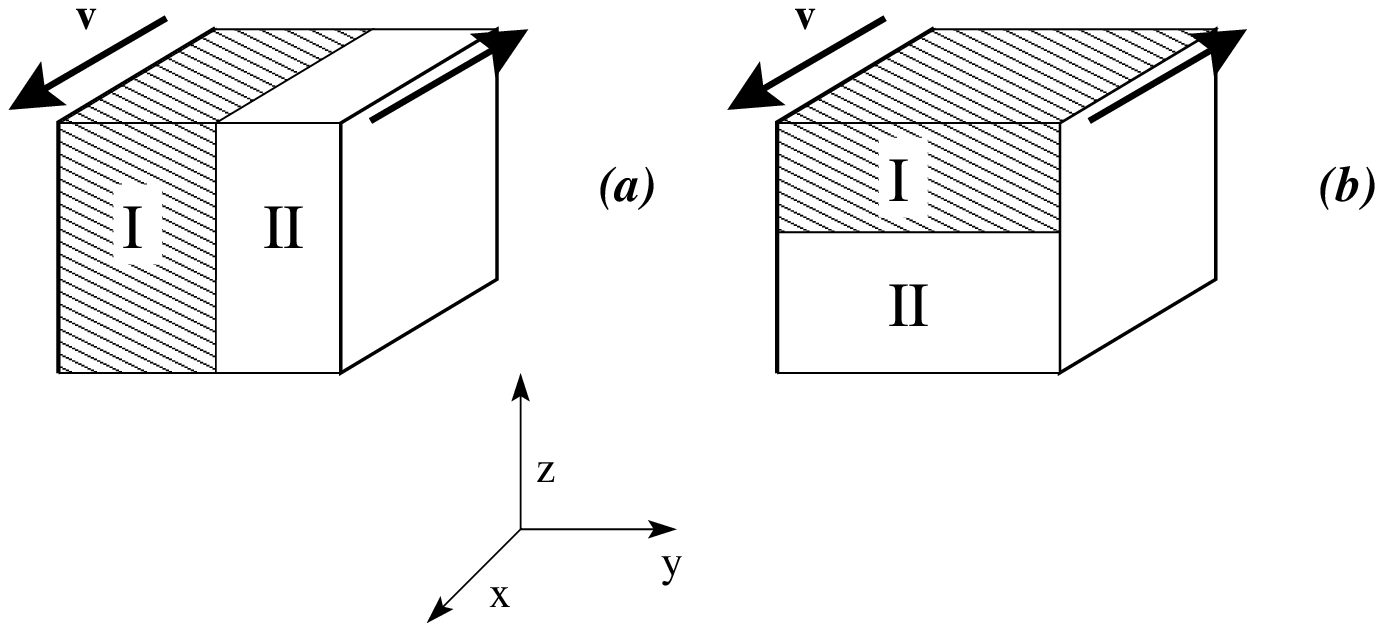}}
\caption{a) Gradient banding: flow-induced phases lie
  in the direction of the velocity gradient (arrow shows flow
  direction). The bands share the same shear stress, but can have
  different shear rates. b) Vorticity banding: bands lie in the flow
  vorticity direction. Here, the shear rate is uniform across the
  bands, but the shear stress can vary from one band to the other.}
\label{banding figure}
\end{figure}

In steady-state there can be no acceleration, so the total stress must
be divergence free.  In planar shear flow, this implies that the shear
rate in the vorticity direction and the shear stress in the gradient
direction are uniform.  Vorticity banding thus corresponds to a
scenario where bands share a common shear rate but can have different
shear stresses (see Figure \ref{banding figure}).  Similarly, when
bands lie in the gradient direction the stress is uniform across the
bands and the shear rate can vary.  Most experiments where banding has
been observed have been carried out in the curved geometries of
cone-and-plate or Couette rheometers.  The gaps in these rheometers
are usually very thin, and in this limit the flow is approximately a
planar shear flow.  [We also note that we consider flows in the low
Reynolds number limit.]

The microscopic mechanisms causing the transitions in all these
complex fluids are likely to be highly system specific, and govern the
critical shear rate or shear stress,
the structure of the flow-induced phases, and the detailed shape of
the flow curves.  At a macroscopic level, however, there appears to be
a high degree of universality between systems. As we have discussed,
different complex fluids can produce qualitatively similar flow
curves. By simply analysing the shape of these flow curves, we have
extracted information about the banding orientation of the transition
\cite{Peter}, as well as the stability of the system \cite{Goveas and
  Pine}.

The obvious analog to this way of thinking is the well-known
Landau-Ginzburg theory of equilibrium phase transitions. A free energy
functional consisting of a double-well local free energy and a square
gradient term reproduces all the phenomenology of a phase transition
in the region of the critical temperature, for many different systems.
However, a microscopic theory is required to calculate the Landau
coefficients.  In this work, we use a multivalued reaction diffusion
scheme to construct a general phenomenological theory to describe
phase transitions in flow; in the spirit of Landau-Ginzburg
theory, such a model could in principle be derived from microscopic theories.

The most difficult step in devising such a nonequilibrium theory is
determining the important variables. Typically there are three
significant quantities: i) a species concentration, which is a
conserved scalar, ii) the momentum density, which is a conserved
vector and iii) the stress, which is a non-conserved tensor.  The
momentum density is described by the Navier-Stokes equation, and its
current is the stress.  In addition non-conserved ``microstructural''
order parameters exist which contribute to the stress. These may be
scalars such as chain length in wormlike micelle solutions, or tensors
such as molecule orientation.  Since all of these variables have
different relative relaxation times, one must distinguish between slow
variables, which require their own equations of motion, and fast
variables, which relax quickly to a steady-state value.

The choice of slow variables affects the structure of the equations of
motion and the couplings between them, and therefore the dynamics of
the system. In models of hydrodynamic instabilities for example, the
momentum is considered to be a slow variable.  A phenomenological
stress constitutive equation is often used: if the stress is taken to
be a fast variable, this relation is simply an algebraic function of
the rate of strain tensor, such as the Newtonian relation for simple
fluids; if the stress is taken to be a slow variable, this relation
takes the form of a differential equation, such as the Upper
  Convected/Oldroyd-B Maxwell model for polymer melts.
The hallmark of complex fluid rheology, however, is the
coupling between the velocity and/or the stress to the microstructure
of the fluid.  In microscopic theories, generally an equation of
motion is not written for the total stress, but for another slow
variable which makes an important contribution to it, such as the
director in nematic liquid crystals or the second moment of the
configuration tensor in polymer melts.

Schmitt, Marques and Lequeux \cite{SchmittMarquesLequeux} have
classified flow instabilities in complex fluids as ``mechanical'' or
``spinodal'' instabilities, using a model where concentration and
momentum are the slow variables.  If a perturbation to the shear rate
first makes the system go unstable, the instability is mechanical,
while it is spinodal if the concentration becomes unstable first.
Note that ``instability'' as it is used here refers to a linear
instability.  Any instability, linear or nonlinear (we return to this
issue at the end of the paper), can lead to a macroscopically
shear-banded state that resolves the instability.

Shear-banding associated with
momentum instabilities have been analyzed in detail at a
high (macroscopic) level, using the phenomenological Johnson-Segalman
model \cite{Spenley,Yuan,Olmsted JS}.  Here a non-conserved
``polymer'' stress tensor, playing the role of the slow variable, is
included in the total momentum density, resulting in a multi-valued
stress constitutive relation. This model produces gradient banding and
a flow curve with a stress plateau, and is considered a reasonable
mimic of shear-thinning wormlike micelles.  Microscopically derived
theories for wormlike micelles \cite{Cates} and nematic liquid
crystalline melts \cite{Olmsted LC} yield a non-monotonic relation
similar to the Johnson-Segalman model, but with the benefit of a
molecular interpretation.

An alternative caricature to these models has been developed in
phenomenological theories for shear-thickening.  Originally, Ajdari
\cite{Ajdari} proposed an equation of motion for the position of an
interface that separates high and low viscosity phases under shear.
By coupling this equation with conservation laws and a Newtonian
stress constitutive equation for the micellar solution, a
non-monotonic flow curve was produced. Goveas and Pine adopted this
approach to describe shear-thickening wormlike micelles and were able
to successfully reproduce much of the experimental phenomenology.  The
flow curve was then used to explain the differences in stress versus
shear rate control, based on a linear stability analysis of the
interfacial height equation.  In this case, the momentum and micellar
solution stress were taken to be fast variables, while an equation of
motion was written for a scalar variable, which is the macroscopic
manifestation of changes in the fluid microstructure.

However, the formulation of Goveas and Pine did not contain any
mechanism for the formation of the shear-induced state, so that the
\textit{existence} of the new phase was simply postulated by the
presence of an interface.  In this paper we present a generic
phenomenological model that naturally admits a flow-induced phase and
incorporates spatial gradients so that an interface structure and its
stability can be determined. The model consists of an equation of
motion for the volume fraction of a reacting species, \textit{i.e.} a
scalar non-conserved order parameter representing microstructural
change in a complex fluid. There are fast stress variables associated
with the reactant and product species, which contribute additively to
the total stress.  We have continuously evolved the model from a
homogeneous to a phase-separated state, and examined how thinning and
thickening flow curves, as well as the size of the gradient terms,
affect phase transitions, and in particular the
banding orientation (vorticity versus gradient banding).  Most
significantly, we are able to probe the nonlinear dynamic behavior of
the system.
\section{Minimal model}
Our phenomenological theory consists of a general reaction-diffusion
scheme. The reaction terms represent the creation and destruction of a
variable under flow, and are analogous to the local free energy terms
in a Landau-Ginzburg theory. This variable may embody a species
concentration, or a structural parameter such as aggregate size or
molecule orientation.  While this scheme is meant to be quite general
and is a vehicle for capturing the general physics for many complex
fluids, a reaction diffusion scheme has a literal basis for wormlike
micelles and onion solutions.  In the wormlike micelle case, such
``reaction'' terms correspond to the constant breaking and
recombination of the ``living'' polymers; while in onion solutions,
the reaction terms might correspond to the formation of onions.  The
steady-state onion size scales as the inverse square root of the shear
rate \cite{Roux} and is a reversible function of the shear rate;
\textit{i.e.}  the size is independent of whether smaller onions are
created by increasing the shear rate applied to larger onions, or
larger onions are created by decreasing the shear rate applied to
smaller onions. This indicates that onion combination and fracture
processes compete to attain steady state, and these processes have
different dependences on shear rate.

In the same way that a double well potential signals the possibility
of equilibrium phase coexistence, a multivalued reaction term can
allow for flow-induced banding.  The diffusion terms are the analog of
the non-local terms in the free energy and provide gradients which can
support inhomogeneities and describe interfaces between states.
However, unlike in equilibrium, where a global minimization
  principle applies, the diffusion terms are necessary for
determining the conditions for phase coexistence in flow
\cite{nonlocal stress}.

Consider a system which is one-phase at equilibrium, and
consists solely of a species, $A$. Planar shear flow is then applied to this system:
the coordinate system is
shown in Figure \ref{banding figure}, where $x,y$ and $z$ denote the
flow, gradient and vorticity directions respectively. We consider only
variations in $y$ and $z$ in this work.  Suppose that a new phase,
$B$, can be induced by flow, such that at a given shear rate (or
stress) a dynamic equilibrium between $A$ and $B$ is established.  We
write this schematically as:
\begin{equation}
  A \rightleftharpoons B.
\end{equation}
Let us define an order parameter, $\phi_{B}\equiv \phi$, which
corresponds to the volume fraction of the $B$-species (although
allowing $\phi$ to correspond to a structural variable is also
viable).  The system is constrained to have constant density such that
\begin{equation}
  \label{density constraint}
  \phi_{A}+\phi_{B}=1.
\end{equation}
Notice that $\phi_{A}$ and $\phi_{B}$ are non-conserved variables,
although the total density is conserved. 
We write the following equation of motion for $\phi$ as
\begin{equation}
\label{deq}
\frac{d \phi}{dt}= R(\phi, \dot{\gamma})+D \nabla^{2} \phi,
\end{equation}
where $R(\phi, \dot{\gamma})$ represents the forward and backward
``reactions'' which create and destroy the new phase,
$\dot{\gamma}(y,z)=dv_x/dy$ is the local shear rate (where $v_{x}$ is
the component of the velocity in the flow direction), and $D$ is an
effective diffusion
coefficient (taken to be a constant). The {\it homogeneous}
steady-state solutions to equation~(\ref{deq}), where $\nabla^{2}
\phi=0$, are given by
\begin{equation}
  \label{reaction eq}
  R(\phi,\dot{\gamma})=0.
\end{equation}

\section{Constitutive curves}\label{flow curve section}
To compute flow curves for the system, stress constitutive equations
for the different components must also be specified. The simplest
possible scheme involves additive Newtonian relations for each species
\begin{equation}\label{stress}
  \begin{split}
    \sigma&=\sigma_{A}+\sigma_{B} \\
    \sigma_{\alpha}&=\eta_{\alpha} \phi_{\alpha} \dot{\gamma},
  \end{split}
\end{equation}
where $\sigma$ is the total shear stress, and $\sigma_{\alpha}$ and
$\eta_{\alpha}$ are the shear stress and viscosity, respectively, of
species $\alpha=A,B$. Stress and composition are thus effectively
coupled in the system. Applying the density constraint, equation
(\ref{density constraint}), gives an expression for the constitutive
curve of the system,
\begin{equation}
\label{constitutive stress}
\sigma=[\phi(c-1)+1] \dot{\gamma},
\end{equation}
where $\phi=\phi(\dot{\gamma})$ is the solution to equation
(\ref{reaction eq}), $c=\eta_{B}/\eta_{A}$ is the ratio of the
viscosities of the two components, and we have set $\eta_{A}=1$ for
simplicity.
\begin{figure}
\epsfxsize=3.3in
\centerline{\epsfbox{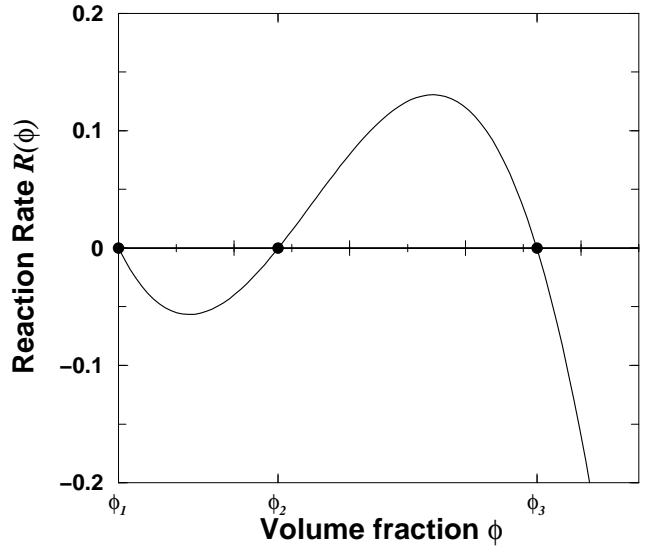}}
\caption{Multivalued reaction scheme
(see equation \ref{deq})
  produc\textbf{ing} three possible roots, $\phi_{1}, \phi_{2}$ and
$\phi_{3}$, corresponding to stable, unstable and stable homogeneous
states, respectively, for a fixed value of the shear rate.}
\label{mfig}
\end{figure}

Notice that a multivalued $R(\phi,\dot{\gamma})$ produces
multiple homogeneous steady-states. Choosing
\begin{equation}
\label{reaction term}
R(\phi,\dot{\gamma})=|\dot{\gamma}| \phi_{A} \phi_{B}^{2}- k \phi_{B} 
\end{equation}
yields the curve shown in Figure \ref{mfig}, for an imposed
$\dot{\gamma}$.  [$k$ represents a rate constant for a backward
reaction, which has dimensions of inverse time and is henceforth set
to unity.]  Notice that the forward reaction term in
equation~(\ref{reaction term}) has a linear shear rate dependence, so
we must take its absolute value from symmetry considerations.  In
microscopic theories for flow-induced reactions in wormlike micelles
\cite{Cates and Turner} and polymers \cite{Glenn}, such an {\it
  effective} reaction rate also has a linear or non-analytic form,
resulting from the projection of tensorial degrees of freedom onto a
scalar order parameter.  For a given local shear rate, the reaction
scheme of equations (\ref{density constraint}, \ref{reaction eq},
\ref{reaction term}) yields the following homogeneous steady-states:
\begin{subequations}
\label{shear steady states}
\begin{align}
  \phi_{1}&=0\\
  \phi_{2}& =\frac{1}{2}- \frac{1}{2}
  \sqrt{1-\frac{4}{\dot{\gamma}}}\\
  \phi_{3}&=\frac{1}{2}+\frac{1}{2} \sqrt{1-\frac{4}{\dot{\gamma}}} .
\end{align}
\end{subequations}
Performing a linear stability analysis on equation~(\ref{deq}), at
fixed $\dot{\gamma}$, shows that $\phi_{1}$ and $\phi_{3}$ are stable
fixed points for the system, while $\phi_{2}$ is an unstable fixed
point.  This is evident from Figure \ref{mfig}, where the middle root
has positive slope $dR/d\phi>0$.

\begin{figure}
\includegraphics[scale=0.5]{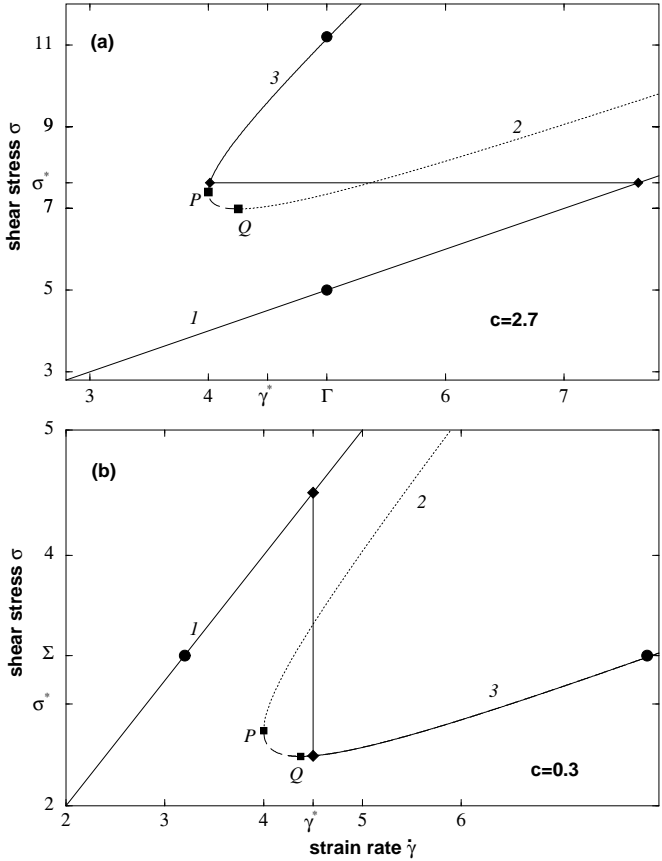}
\caption{Flow curves (from equations \ref{reaction eq},
\ref{constitutive stress} and  \ref{reaction term}) corresponding to
  the minimal model.  At low shear rates, the flow curve is single-valued
  (branch $1$). Above a certain shear rate (or stress),
an additional unstable branch, $2$, and a stable
  branch, $3$, exist. These branches correspond to the homogeneous roots
$\{\phi_1,\phi_2, \phi_3 \}$ from equations~(\ref{shear steady states})
for imposed shear rate, and $\{ \phi_1,\phi_2', \phi_3' \}$ from 
equations~(\ref{phi stress}) for imposed stress. In the former case,
branches $2$ and $3$ are separated by $P$, while they are
separated by $Q$ in the latter. Thus, the line segment between $P$ and $Q$ 
is stable under
  controlled shear rate, but unstable under controlled stress.
(a) Shear-thickening flow curve for $c>1$, illustrating the
    controlled shear rate case. At a fixed shear rate, $\Gamma$, the
    system can choose between homogeneous states on branches $1$ and
   $3$, or gradient band between these branches at stress
   $\sigma^{*}$.  (b) Shear-thinning flow curve for $c<1$,
    illustrating the constrolled stress case.  At fixed stress
  $\Sigma$, the system can vorticity band at shear rate
   $\dot{\gamma}^{*}$ between $1$ and $3$, or choose between
   homogeneous states on these same branches.}
\label{minimal flow curves}
\end{figure}
Substituting the homogeneous roots from equation~(\ref{shear steady
  states}) into equation~(\ref{constitutive stress}) produces three
branches of the constitutive curve (labelled as $1, 2$ and $3$
respectively in Figure \ref{minimal flow curves}).  Notice that below
a certain shear rate, $\dot{\gamma}=4$, only the $\phi_{1}$ root is
real and the reaction curve is single-valued. This is marked as point
$P$ on Figure \ref{minimal flow curves}, and has coordinates
\begin{equation}
  \left\{\dot{\gamma}_{P},\sigma_P\right\} =  \left\{4,
    2(c+1) \right\}. 
\end{equation}
Physically, this means that only
species $A$ exists at low (uniform) shear rates.  From
equation~(\ref{constitutive stress}), we can see that this one-phase
system ($\phi_{A}=1$) is Newtonian, and the corresponding flow curve
has a slope of unity.  The slope of the stable flow-induced branch $3$
depends on the value of the parameter $c$.  For $c<1$, a transition
from branch $1$ to branch $3$ is shear-thinning, while for $c>1$ such
a transition is shear-thickening. The locus of flow induced roots
$(\phi_2,\phi_3)$ exhibits a minimum in
the shear stress as a function of shear rate, which is denoted as
point $Q$ in Figure \ref{minimal flow curves}: 
\begin{equation}
  \label{stressmin}
  \left\{\dot{\gamma}_{Q},\sigma_Q\right\}=
  \left\{\frac{(1+\sqrt{c})^2}{\sqrt{c}}, (1+\sqrt{c})^2 \right\}.
\end{equation}

For a given value of the local stress, the homogeneous steady-states
are given by
\begin{subequations}
\label{phi stress}
\begin{align}
  \phi_{1}& =0 \\
  \phi_{2'}& =\frac{ (\sigma-c+1)-
    \sqrt{(\sigma-c+1)^{2} - 4 \sigma}} {2 \sigma}\\
  \phi_{3'}& =\frac{ (\sigma-c+1)+ \sqrt{(\sigma-c+1)^{2} - 4 \sigma}}
  {2 \sigma}.
\end{align}
\end{subequations}
Note that these roots are found by recasting equation~(\ref{reaction
  term}) in terms of the stress, by using equation~(\ref{constitutive
  stress}). This procedure is not equivalent to substituting equation
(\ref{constitutive stress}) into equation~(\ref{shear steady states}).
This is because while the loci of homogeneous states is the same under
fixed local stress or shear rate, the {\it stability} of these
steady-states is not; \textit{i.e.} the portion of the constitutive
curve between $P$ and $Q$ is unstable under fixed local stress, but
stable under fixed local shear rate.  In Figure \ref{minimal flow
  curves}, point $Q$ marks the stress above which the constitutive
curve is multivalued for controlled stress, while $P$ marks the strain
rate above which the constitutive curve is multivalued for controlled
shear rate.

In an experiment, however, only the {\it average} stress and
shear rate can be controlled. If the average shear rate is held fixed
at $\Gamma$, for example, the system can choose between various
options (illustrated in Figure \ref{minimal flow curves}a):
\begin{itemize}
\item[i)]A homogeneous low stress state, $\phi_{1}$.
\item[ii)]A homogeneous high stress state, $\phi_{3'}$.
\item[iii)] A mixture of states (i) and (ii), where the interfaces
  between phases lies in the vorticity direction (vorticity banding).
  Note that (i) and (ii) cannot coexist with each other in the
  $y$-direction, since the stress must be homogeneous in the gradient
  direction.
\item[iv)] A mixture of high shear rate phase, $\phi_{3'}$ and a low
  shear rate phase, $\phi_{1}$. Here the system attains
  an intermediate stress, $\sigma^{*}$, and the
  relative proportions of the two phases are set such that the average
  shear rate is maintained at $\Gamma$.  Since the bands have the same
  stress, but different shear rates, this scenario corresponds to
  gradient banding.
\end{itemize}
While it appears from Figure \ref{minimal flow curves} that there is a
multiplicity of stresses $\sigma^*$
at which the system can gradient band, in fact the system selects a
particular stress (see next Section).

If instead, the average stress is fixed at $\Sigma$ (illustrated in
Figure \ref{minimal flow curves}b), the system can choose a high or
low shear rate homogeneous phase, or it can gradient band between
these. Alternatively, it can band in the vorticity direction between
high and low stress states, by adopting a shear rate
$\gamma^{*}$. There is also a selected shear rate for vorticity
banding.  {\it The essential question is: which of the many possible
  states available to it does the system actually choose and why?}

\section{Calculating the banding stress and shear rate}
The inclusion of gradient terms in equation~(\ref{deq}) causes stress
selection for gradient banding, and shear rate selection for vorticity
banding \cite{nonlocal stress}.  The selected stress and shear rate
are determined by mathematically connecting two different homogeneous
stable states to form an inhomogeneous profile.  To find the banding
shear rate $\dot{\gamma}^{*}$ at which vorticity banding can
  occur, equation~(\ref{deq}) is integrated across the domain at
steady-state. A banding solution (homogeneous phases separated by
interfaces) is, by definition, one which has no gradients in $\phi$ at
the boundaries. We obtain the following condition:
\begin{equation}
\label{banding eq}
\int_{\phi_{1}(\dot{\gamma}^{*})}^{\phi_{3}(\dot{\gamma}^{*})} d \phi
 \, R[\phi, \dot{\gamma}^{*}]=0.
\end{equation}
where $\phi_{1}$ and $\phi_{3}$ are given by equations~(\ref{shear
  steady states}).
Defining a new function, $F[\phi,\dot{\gamma}]=\int_{0}^{\phi} d \,
\phi' R[\phi', \dot{\gamma}]$, we can rewrite equation~(\ref{banding
  eq}) as
\begin{equation}
  \label{bandq}
  F[\phi_{1}(\dot{\gamma}^{*}),\dot{\gamma}^*]=F[\phi_{3}
  (\dot{\gamma}^{*}),\dot{\gamma}^*], 
\end{equation}
which is analogous to the common tangent construction from equilibrium
thermodynamics. If this were an equilibrium system, $F$ would be
identified as the free energy. Since this is a dynamic system however,
$F$ cannot be given the same physical interpretation, so that the
analogy is purely formal.  Using equation~(\ref{shear steady states})
in equation~(\ref{bandq}) gives
\begin{equation}
\dot{\gamma}^{*}=4.5.
\end{equation}

To calculate the banding stress $\sigma^{*}$ at which gradient
  banding can occur, equation~(\ref{banding eq}) must be recast in
terms of the shear stress using the stress constitutive relation,
equation~(\ref{constitutive stress}) to obtain a relation
  $\dot{\gamma}(\sigma,\phi)$. The banding stress in our minimal
model is only a function of $c$ and is given by the solution of the
following equation:
\begin{equation}
  F[\phi_{1'}(\sigma^{*}),\sigma^*]=F[\phi_{3'}(\sigma^{*}),\sigma^*],
\end{equation}
where $F[\phi,\sigma^*]=\int_{0}^{\phi} d \, \phi'
  R[\phi', \dot{\gamma}(\sigma,\phi')]$ yields
\begin{align}
    F[\phi(\sigma^{*}), \sigma^{*}]&=\frac{\sigma^{*}}{(c-1)^{4}} \left
    [ -\tfrac13\tilde{\phi}^{3} +
    \tfrac12(c+2)\tilde{\phi}^{2} + \tfrac32 c\right.\label{eq:Fbig}\\
  &\left.\phantom{\tfrac13} -(2c+1) \tilde{\phi}  +
    \tfrac13 + c \ln
    \tilde{\phi} \right ] - \tfrac12\phi^{2}(\sigma^*), \nonumber\\
    \tilde{\phi}&=\phi(\sigma^{*})(c-1)+1,
\end{align}
and $\phi_{1'}$ and $\phi_{3'}$ are given by equations~(\ref{phi
  stress}).  The selected stress $\sigma^*$ and shear rate
$\dot{\gamma}^*$ are shown in Figure \ref{minimal flow curves}
for $c=0.3, 2.7$, and given in Table~\ref{tab:bandingvalues}.
\begin{table}[htbp]
  \begin{center}
    \begin{tabular}{|ccccccc|}
      \hline
      $c$ & $\sigma^{\ast}$ & $\phi_3(\sigma^{\ast})$  & 
      $\dot{\gamma}(\phi_3^{\ast})$ &$\sigma_P$& $\dot{\gamma}_Q$
      &$\sigma_Q$\\\hline 
      $0.3$ & $2.815$ & $0.810$ & $6.504$& $2.6$&$4.374$ &$2.395$\\
      $0.6$ & $3.606$ & $0.732$ & $5.100$& $3.2$&$4.066$ &$3.149$\\
      $1.2$ & $4.910$ & $0.642$ & $4.352$& $2.4$&$4.008$ &$4.391$\\
      $2.7$ & $7.625$ & $0.529$ & $4.014$& $5.4$&$4.252$ &$6.986$\\\hline
    \end{tabular}
    \caption{Banding stress $\sigma^{\ast}$, points of instability $P$
      and $Q$, and coexistence conditions for different values of $c$.
      In all cases $\dot{\gamma}_P=4$ and $\dot{\gamma}^*=4.5$., while
      the stress and shear rate on branch $\phi_1$ are related by
      $\sigma_1=\dot{\gamma}_1$.}
    \label{tab:bandingvalues}
  \end{center}
\end{table}

\section{Dynamical Selection of Steady-States}
In the preceding Sections we have seen that certain homogeneous and
banded states are available to the system, based on a {\it
  steady-state} analysis.  To determine which of these states is
selected in practice, equation~(\ref{deq}) must be evolved in time; to
make contact with experiments we can only impose constraints of fixed
average shear rate or stress. In this work only one-dimensional
calculations are performed, so that the equation of motion is solved
either in the $y$ (gradient) or $z$ (vorticity) directions.  If there
are composition modulations in the $y$-direction, these can only cause
modulations in the shear rate (gradient banding), since the shear
stress must be uniform in $y$. If the average stress is controlled,
there is only one ``interesting'' stress $\sigma^{*}$ at which the
system can become inhomogeneous in the $y$ direction. Due to
the numerical difficulty of fixing a precise stress in the system, we
do not consider this case.  By comparison, if the average shear rate
is controlled, there is a wide range of shear rates for which
we can investigate whether the system remains homogeneous or gradient
bands.  Similarly, when
composition modulations in $z$ are allowed, we can only look for
vorticity banding under controlled stress within this calculation.
Gradient banding can occur only if the shear rate is set exactly at
$\dot{\gamma}^{*}$, which we do not study here.
\subsection{Controlled Average Shear Rate}\label{shear rate}
We first consider the system under shear rate control, and only allow
for spatial variations in $y$.  Integrating the stress
relation, equation~(\ref{constitutive stress}), across the
domain (where the shear stress is independent of $y$) gives the local
shear rate as a function of $V$, the velocity difference across the
system.  Then equation~(\ref{deq}) becomes
\begin{equation}
\label{average shear rate}
\frac{d \phi}{dt}= \frac{V}{\int dy
  \left\{1/\left[\phi\left(c-1\right) + 1\right]\right\}}
  \frac{\phi^{2}(1-\phi)}{\phi(c-1)+1} 
  -\phi +D \frac{d^{2} \phi}{dy^{2}}.
\end{equation}
This is an integro-differential equation, instead of the differential
equation which yielded the analysis of Section \ref{flow curve
  section}.  Notice, however, that the same homogeneous steady-states
are obtained.  Equation~(\ref{average shear rate}) is solved using
random initial conditions with $\phi$ uniformly chosen within the
range $[0-1]$, and no flux boundary conditions, keeping $V$ at a fixed
value. The domain size is normalized to unity, so that $V$ is
synonymous with the average shear rate.

To solve equations~(\ref{average shear rate}), we use a fully implicit
finite difference scheme, using a central difference approximation for
first and second spatial derivatives, and a forward difference
approximation for the time derivative.  Nonlinear terms are linearized
in time as follows:
\begin{equation}
W[\phi(x,t+\Delta t)]=W[\phi(t)]+ [\phi(t+\Delta t)-\phi(t)]\frac{d
  \,W[\phi,t]}{d \phi}.
\end{equation}
The integral in equation~(\ref{average shear rate}) is evaluated
explictly, \textit{i.e.} at the previous time step. In general, $300$
spatial mesh points are used with a time-step of $1/10000$ \cite{mesh points}.

For some initial conditions, the resulting steady-states are
homogeneous, while for others, they are banded. This implies that
there is a {\it basin of attraction} for attaining a banded state.
Our results can be categorized according to the shape of the
constitutive curve and the magnitude of the diffusion coefficient, and
whether the system is shear-thinning or thickening.  We find that
decreasing the diffusion coefficient increases the basin of attraction
of the banded state and destabilizes the homogeneous state.  That is,
the system is more likely to band for narrow interfaces.  Intuitively,
this makes sense since the banded state represents a mathematical
connection between two homogeneous states: the wider the interface,
the more difficult it is for gradients to be non-zero near the
boundaries of the system.
\begin{figure}
\centering{\includegraphics[scale=0.5]{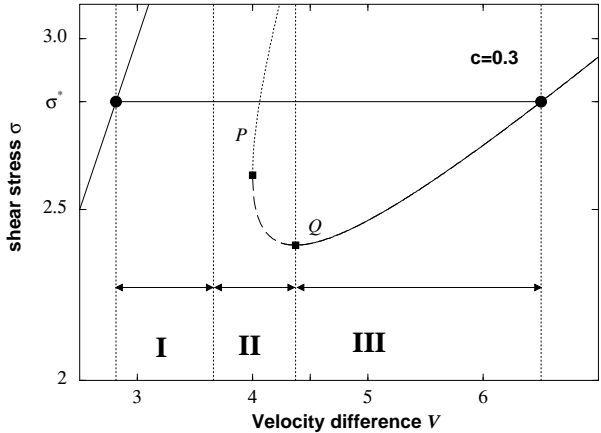}}
\caption{Basins of attraction for different states, for $c=0.3$ and
  imposed mean shear rates $V\equiv\bar{\dot{\gamma}}$. In \textbf{I}
  the system usually attains the homogeneous flow branch
  $\phi=\phi_1=0$. In \textbf{II} the system usually bands at stress
  $\sigma^*$, and in \textbf{III} the system usually attains the
  homogeneous state $\phi=\phi_3$. The behavior is smooth as $V$ is
  increased through these regimes (see data in
  Table~\ref{tab:basins}).}
\label{fig:basinregions}
\end{figure}
For $c<1$, for $V<\dot{\gamma}_{P}$ the system usually chooses branch
1 of the constitutive curve over the banded state, even for small
diffusion coefficients. 
Notice that banding is first allowed, in principle, when the imposed
shear rate is larger than that of the low shear rate band.
The ``critical'' shear rate, where the
system actually first starts to band, is generally somewhat higher than
this:  it is the low shear rate limit of the
stress plateau that would be measured experimentally.
Below the critical shear rate
the system {\it always} chooses branch 1,
although the exact location of this point depends on the diffusion
coefficient. [Increasing the diffusion coefficient widens the
interface, affecting where the interface first ``touches" the wall
\cite{Olmsted JS} and the ability of a banded system to satisfy the
boundary conditions, as discussed above.]

\begin{figure}
\vskip1.0truecm
\epsfxsize=3.0truein
\centering{\epsfbox{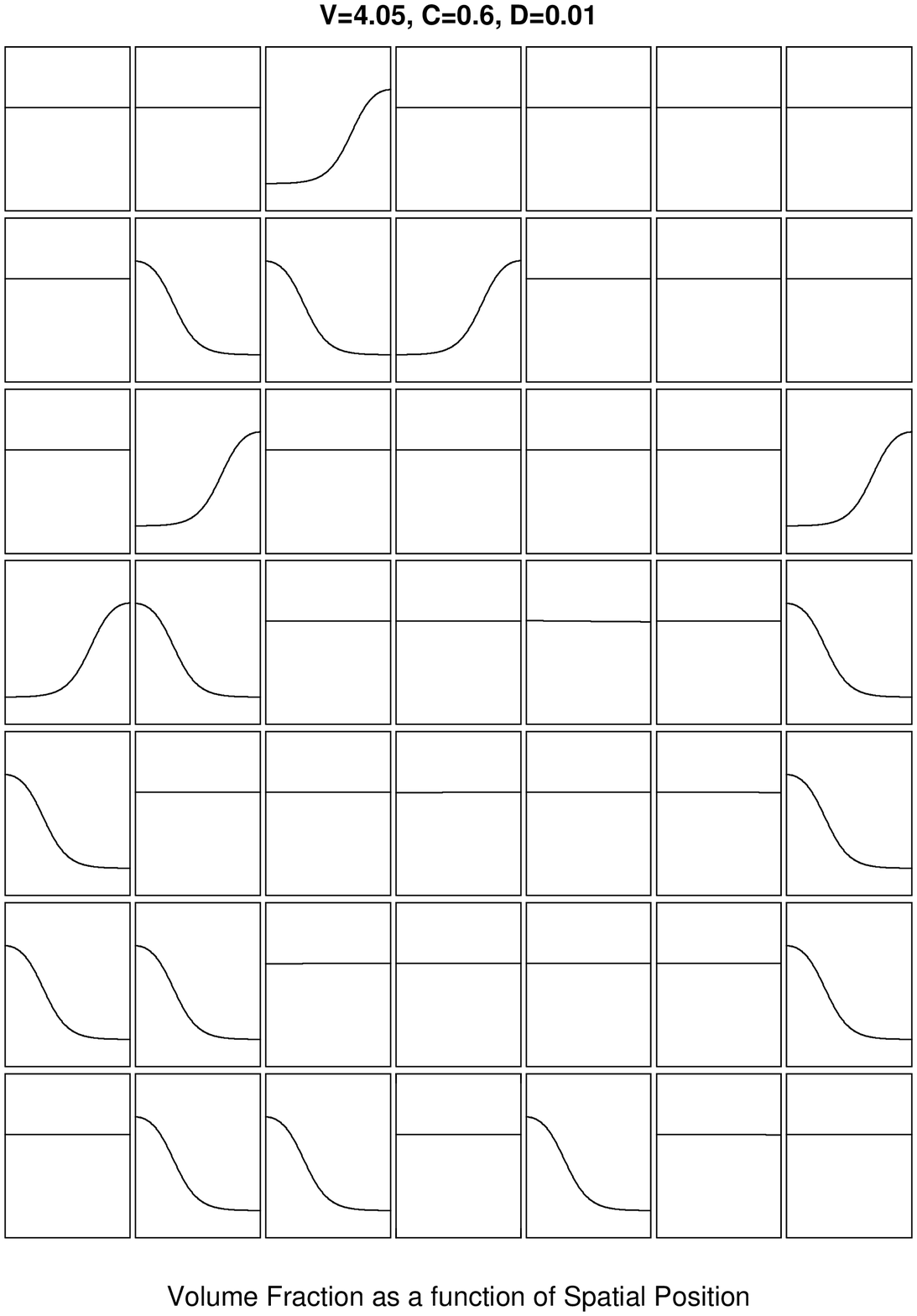}}
\vskip1.0truecm
\epsfxsize=3.0truein
\centering{\epsfbox{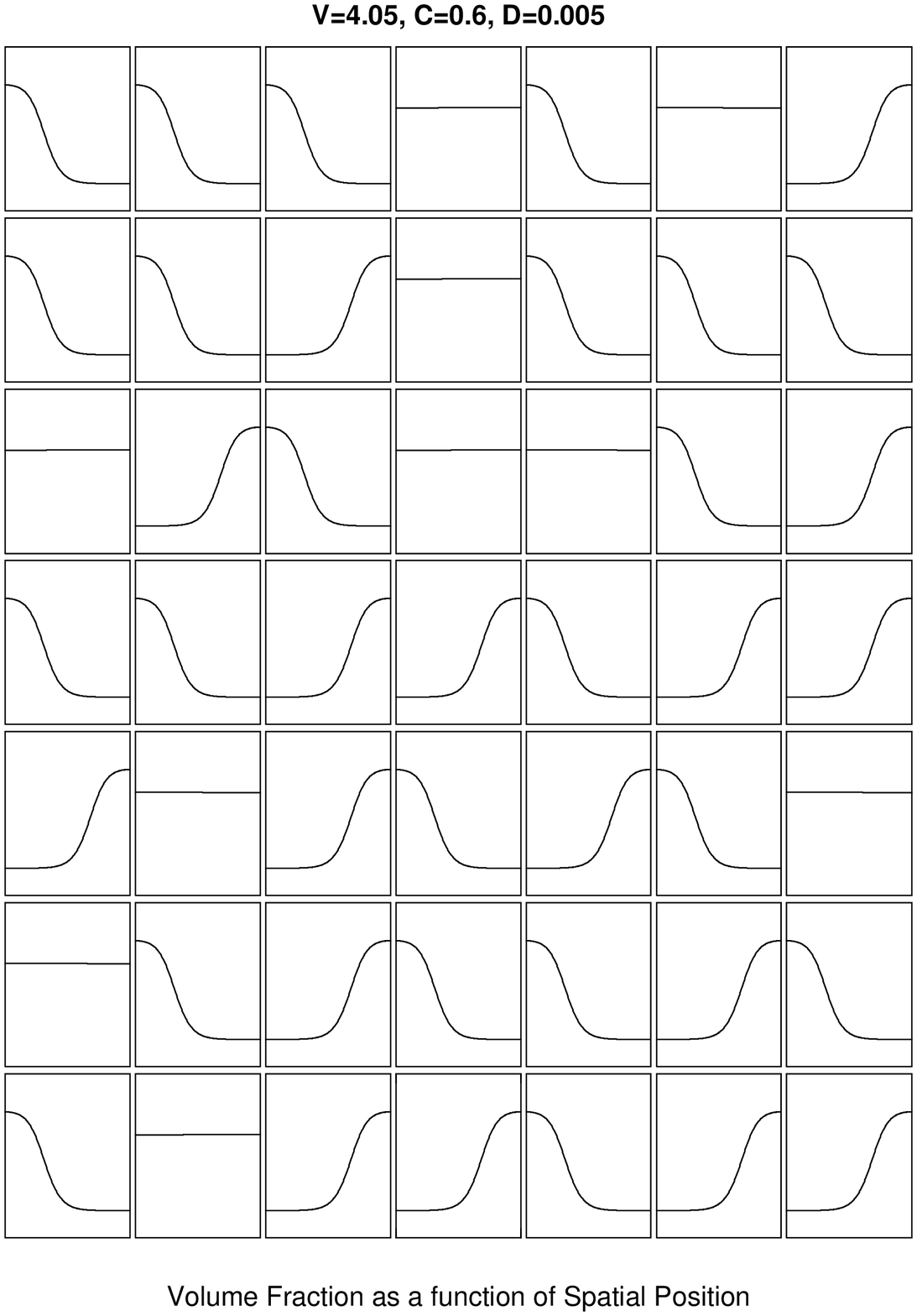}}
\caption{Steady-state composition profiles
  for 49 runs with random initial conditions, for $c=0.6, V=4.05$.
  Here, $\dot{\gamma}_{P}$ corresponds to $V=4.0$, and
  $\dot{\gamma}_{Q}=4.066$.}
\label{zbandingb}
\end{figure}

\begin{figure}
\vskip1.0truecm
\epsfxsize=3.0truein
\centering{\epsfbox{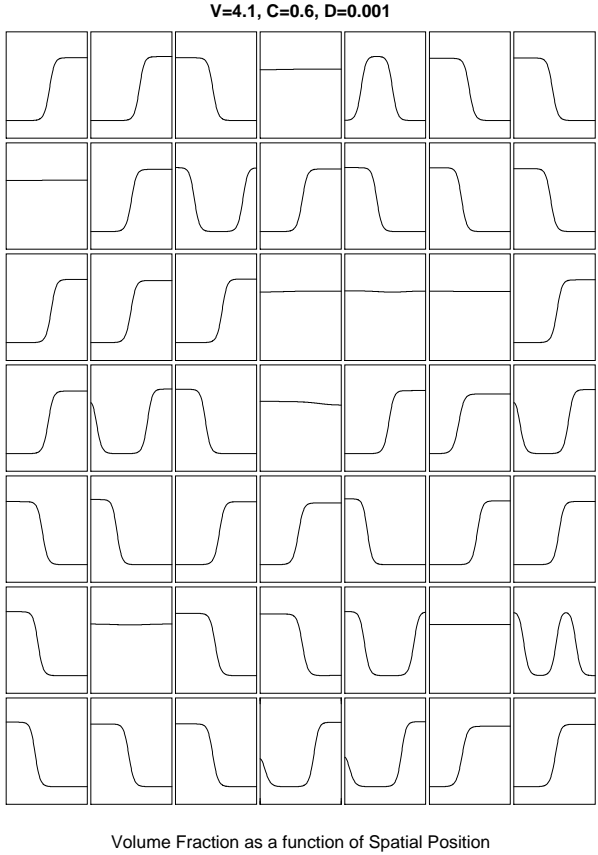}}
\vskip1.0truecm
\epsfxsize=3.0truein
\centering{\epsfbox{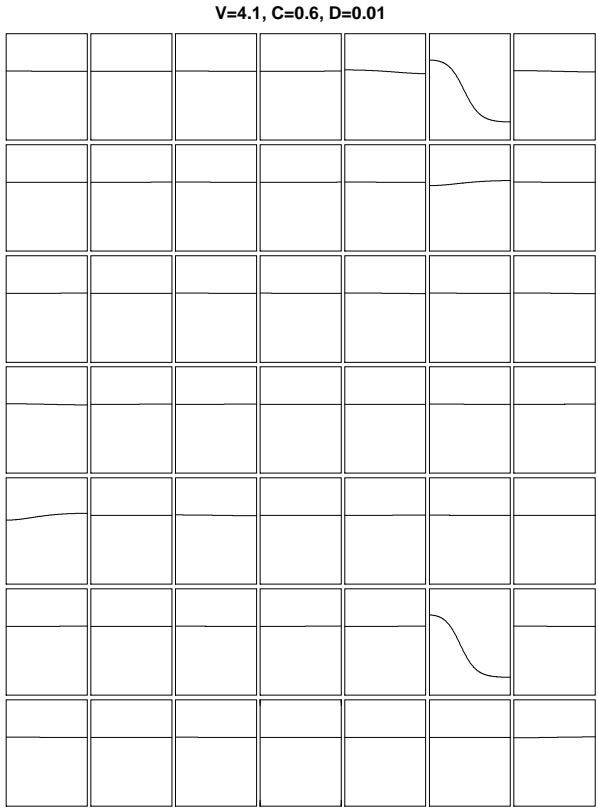}}
\caption{Steady-state composition profiles for 49 runs
with random initial conditions, for $c=0.6, V=4.1$}
\label{zbandinga}
\end{figure}

When $\dot{\gamma}_{P}< V < \dot{\gamma}_{Q}$, the system is more
inclined to band rather than remain homogeneous, even for large
diffusion coefficients (see Fig.~\ref{fig:basinregions}).  The chosen
homogeneous states always lie on branch 3 of the constitutive curve.
Beyond $\dot{\gamma}_{Q}$, there is a crossover to where
homogeneous states (on branch 3) are preferred over banded states.
The exact crossover depends on the value of the diffusion coefficient.
This shows that the system has made a
transition under flow, so that for $c<1$ a shear-thinning transition
is seen, as discussed in Section \ref{flow curve section}.  These
findings are illustrated in Table 2, for 49 runs with different
initial conditions with $c=0.3$.  The ``critical'' shear rate in this
case is around $V=3.3$.  Figures \ref{zbandingb} and \ref{zbandinga}
show similar behavior for $c=0.6$.

Gradient banding is never observed in our numerical experiments when
$c>1$.  Here, the chosen final steady state, above the critical shear
rate, is always the stable high stress homogeneous state on branch 3,
which makes this a shear-thickening transition.

\begin{table}[htbp]
\caption{Summary of results for $c=0.3$, for $49$ runs with random initial
  conditions and controlled mean shear rate $V$. In this case, $\dot{\gamma}_{P}$
corresponds to $V=4.0$, and $\dot{\gamma}_{Q}$ corresponds to $V=4.374$.
Banding is first allowed at $V=2.815$.}
\begin{center}
\begin{tabular}{|l|@{\qquad}l@{\qquad}ccc|}
\hline
 &  & & Result & \\
V & D & $\phi_1$&  banded & $\phi_3$  \\\hline
3.3   & 0.01 & 49 & 0 & 0   \\
& 0.005    &  49 & 0 & 0 \\
& 0.001    &  48 & 1 & 0 \\\hline
3.6   & 0.01 &  45 & 5 & 0  \\
& 0.005    &  34 & 15 & 0 \\
& 0.001    &  6 & 43 & 0 \\\hline
3.8   & 0.01 &  15 & 34 & 0  \\
& 0.005    &  8 & 41 & 0 \\
& 0.001    &  0 & 49 & 0 \\\hline
3.9   & 0.01 & 0 & 49 & 0   \\
& 0.005    & 0 & 49 & 0  \\
& 0.001    & 0 & 49 & 0  \\\hline
4.0   & 0.01 & 0 & 49 & 0   \\
& 0.005    &  0 & 49 & 0 \\
& 0.001    &  0 & 49 & 0 \\\hline
4.05   & 0.01 & 0 & 49 & 0  \\
& 0.005    &0 & 49 & 0  \\
& 0.001    &0 & 49 & 0  \\\hline
4.1   & 0.01 & 0 & 49 & 0  \\
& 0.005    &0 & 49 & 0  \\
& 0.001    &0 & 49 & 0  \\\hline
4.25& 0.01 &  0 & 39 & 10  \\
& 0.005    & 0 & 47 & 2  \\
& 0.001    &0 & 49 & 0  \\\hline
4.37& 0.01 & 0 & 10 & 39  \\
& 0.005    & 0 & 33 & 16  \\
& 0.001    &0 & 49 & 0  \\\hline
4.4& 0.01 & 0 &  10 & 39  \\
& 0.005    & 0 & 21 & 28  \\
& 0.001    &0 & 49 & 0  \\\hline
4.55& 0.01 & 0 &  2 & 47  \\
& 0.005    & 0 & 8 & 41  \\
& 0.001    & 0 & 13 & 36  \\\hline
4.6& 0.01 & 0 &  1 & 48  \\
& 0.005    & 0 & 7 & 42  \\
& 0.001    & 0 & 20 & 29  \\\hline
4.7& 0.01 &  0 & 0 & 49  \\
& 0.005    & 0 & 4 & 45  \\
& 0.001    & 0 & 19 & 30  \\\hline
\end{tabular}
\end{center}
\label{tab:basins}
\end{table}
The preceding results are for the case where the shear rate is held at
a steady value, and might apply to a system where the mean shear rate
is applied to an initially noisy system. However, most experiments are
conducted by starting up the system from zero shear rate, and then
discontinuously ramping the shear rate to higher values.  We have
tried to mimic this scenario, by bringing the system to steady-state
for $V<\dot{\gamma}_{P}$, and then suddenly increasing $V$ to greater
than $\dot{\gamma}_{P}$.  In order to dislodge the system from branch
1 to the banded state, however, we need to add noise of amplitude
order unity.  Such a large amount of noise essentially obliterates any
memory of the initial steady-state, suggesting that a nucleation event
is required for an experimental system to
band from start-up, as was found previously in Ref.~\cite{Olmsted JS}.


In general, the banded state consists of two bands, corresponding to
branches 1 and 3 on the constitutive curve.  For some initial
conditions, multiple bands are found.  We do not attach much
significance to this result, because it is known that multiple
interfaces are allowed for planar flow, and the number of allowed
interfaces in such one-dimensional systems is known to increase as the
diffusion coefficient decreases \cite{Grindrod book}.  Britton and
Callaghan have reported multiple gradient bands for wormlike micelles
in Couette flow \cite{Callaghan}. However, it has been shown
\cite{Olmsted JS} that simple constitutive relations (like the one
derived here) do not permit multiple interfaces in Couette flow. This
implies that the current model cannot describe these experimental
observations.
\subsection{Controlled Average Stress}
Next we fix the average shear stress and solve the model in the
$z$-direction, hence allowing for vorticity banding at
  different stresses. Following a procedure analogous to that of the
previous Section, the observation that the shear rate is uniform in
$z$ allows us to convert equation~(\ref{deq}) into the following
integro-differential equation
\begin{equation}
\label{average stress}
\frac{d \phi}{dt}= \frac{\Sigma  \,  \phi^{2}(1-\phi)}{\int dz
  [\phi(c-1)+1]} -\phi +D \frac{d^{2} \phi}{dz^{2}},
\end{equation}
where $\Sigma$ is the imposed average shear stress across the
domain.
 
Note that interchanging the stress and shear rate variables in the
phenomenological model would ``reverse'' all the results of Section
\ref{shear rate}.  Thus, branch 3 would be thickening for $c<1$, and
the system would be inclined to vorticity band under stress control,
for $\sigma_{P}<\Sigma<\sigma_{Q}$. Similarly, the system would remain
homogeneous for $c<1$.  This reasoning implies that the shape of the
flow curve sets the attractors for the system. If this is true, then
when stress and shear rate are not interchanged, the system should be
predisposed to vorticity band for $c>1$ when the stress is fixed between
$\sigma_{P}$ and $\sigma_{Q}$, but should remain homogeneous for 
all stresses when $c<1$.

This is indeed what we find from numerical solution of
equation~(\ref{average stress}). Table~\ref{tab:basinsig} shows
results for various values of $\Sigma$ and $c=2.7$. Decreasing the
diffusion coefficient destabilizes the homogeneous state, as in the
controlled shear rate case.
\begin{table}[htbp]
\caption{Summary of results for $c=2.7$, for $49$ runs with random initial
  conditions and controlled stress $\Sigma$. In this case, $\sigma_{P}=
5.4$ and $\sigma_{Q}=6.896$.}
\begin{center}
\begin{tabular}{|l|@{\qquad}l@{\qquad}ccc|}
\hline
 &  & & Result & \\
$\Sigma$ & D & $\phi_1$&  banded & $\phi_3$  \\\hline
5.0   & 0.01 &  49 & 0 & 0  \\
& 0.005    &  49 & 0 & 0 \\
& 0.001    &  49 & 0 & 0 \\\hline
5.4   & 0.01 &  49 & 0 & 0  \\
& 0.005    &  49 & 0 & 0 \\
& 0.001    &  49 & 0 & 0 \\\hline
6.0   & 0.01 &  49 & 0 & 0  \\
& 0.005    &  48 & 1 &0 \\
& 0.001    &  36 & 13 &0 \\\hline
7.1   & 0.01 & 28 &  21 &0  \\
& 0.005    &  0 & 49 & 0 \\
& 0.001    &  0 & 49 & 0 \\\hline
7.2   & 0.01 & 0 &  4 &45  \\
& 0.005    &  0 & 40 &9 \\
& 0.001    &  0 & 49 & 0 \\\hline
7.3   & 0.01 & 0 & 0 & 49  \\
& 0.005    &  0 & 9 &40 \\
& 0.001    &  0 & 43 &6 \\\hline
7.4   & 0.01 &  0 & 0 &49  \\
& 0.005    &  0 & 2 &47  \\
& 0.001    &  0 & 23 &26  \\\hline
7.5   & 0.01 &  0 & 0 &49  \\
& 0.005    &  0 & 0 &49 \\
& 0.001    &  0 & 13 &36  \\\hline
8.5   & 0.01 &  0 & 0 &49  \\
& 0.005    &  0 & 0 &49 \\
& 0.001    &  0 & 0 &49 \\\hline
\end{tabular}
\end{center}
\label{tab:basinsig}
\end{table}

We sometimes observe multiple bands under stress control, as with the
fixed shear rate cases.  Bonn \textit{et al.}  \cite{Bonn} and Chen
\textit{et al.} have seen multiple bands in the vorticity direction in
Couette flow which is probably due to a combination of the slow
coarsening expected in one-dimensional systems \cite{OneD} and
multiple allowed interfaces \cite{Grindrod book}. The stress is
non-uniform and monotonic in the flow gradient direction of
a cylindrical Couette device, which implies a
single stable interface.  The cylindrical geometry does not, however,
impose such an inhomogeneity along the vorticity direction.
\section{Conclusions}
We have shown that a simple phenomenological reaction-diffusion scheme
can produce a flow-induced phase transition, as a consequence of a
multi-valued reaction term. The model consists of an equation of
motion for a non-conserved composition variable, while the stresses
induced in the reactants and products are assumed to be fast
variables. The character of the model depends on a single parameter
$c$, that controls whether or not the transition is shear-thinning or
shear-thickening. Above a critical
shear rate (or shear stress), the system may band or remain
homogeneous.  The steady-states that are selected from random initial
conditions depend on the shape of the constitutive curves and the
magnitude of the diffusion coefficient:

\begin{enumerate}
\item \textit{Imposed shear rates:} For $c<1$ (shear-thinning
  transition), the system chooses a low stress homogeneous state at
  low shear rates.  Above a critical shear rate, gradient
  banding tends to occur for imposed shear rates around the region of
  the constitutive curve with negative slope $d\sigma/d\dot{\gamma}<0$
  (see Figure \ref{minimal flow curves}a).  At shear rates higher than
  this, the system is predisposed towards the high stress homogeneous
  state. For $c>1$ (shear-thickening transition), the system always
  chooses this homogeneous state above the critical shear
  rate and gradient banding is never observed.
\item \textit{Imposed stress:} For $c>1$ (shear-thickening
  transition), the system chooses a low shear rate homogeneous at low
  stresses.  Above a critical stress, vorticity banding tends
  to occur for imposed stresses around the region of the constitutive
  curve with negative slope. For higher stresses, the system is
  predisposed towards the high shear rate homogeneous state.  For
  $c<1$, the system always chooses this homogeneous state above the
  critical stress and vorticity banding is never observed.
\item In the regions of the flow curve where banding is observed, we
  find the apparent basin of attraction for banding increases upon
  decreasing the value of the diffusion coefficient.
\end{enumerate}
While banding is more pronounced in the vicinity of the flow curve
with a negative slope (where the system is linearly unstable), it is
also observed in regions of the flow curve with positive slope. In
particular, the critical shear
rate or stress (where banding is first initiated), lies in the latter
section of the flow curve. Here, the system is nonlinearly unstable to
perturbations. Such behavior has been seen in experiments on
shear-thinning wormlike micelles \cite{grand}, where the onset of
banding occurs at a lower stress (and shear rate) if the system is
given enough time to explore all fluctuations, as compared to where
banding is induced upon rapidly varying the control parameters.  Porte
\textit{et al.} \cite{PorteBerretHarden} have discussed various flow
curves which can contain both linearly and nonlinearly unstable
regions: the equilibrium analog of the former is the spinodal curve,
and that of the latter is the metastable region, where an instability
must be nucleated.

Our results are significant because they
show that a minimal model can exhibit a rich phenomenology, and that
the selection rules for phase coexistence are simple.  To understand
why the system chooses certain states over others in some regions of
the flow curve, a nonlinear dynamics analysis of the model must be
performed.  We believe that our scheme represents a new class of
reaction-diffusion equations, because the constraint of fixed average
stress or shear rate turns the governing partial differential
equations into integro-differential equations, which represents a
general class of dynamical equations that, to our knowledge, has not
been studied.  This system exhibits fascinating and complex nonlinear
dynamics, which we will discuss in a future publication.

Our current scheme is missing much physics: a complete model would
involve coupled equations of motion for conserved variables
(concentration of the various species) and non-conserved, tensorial
variables (structural variables, stress).  Also, we have assumed that
the individual species obey Newtonian stress constitutive relations.
Typically, these species are themselves complex fluids, and
are either shear-thinning or exhibit a yield stress.  In future work,
by systematic exclusion of certain dynamic variables, we will be able
to investigate the individual roles played by the stress,
concentration etc., in order to determine which variables are
essential to the problem formulation.

One of us \cite{Olmsted LC} has already considered a theory with
stress, momentum and concentration variables in the context of rigid
rod suspensions. Separate phase diagrams for shear-induced phase
separation in both the vorticity and gradient directions were
calculated, but the model was too prohibitively complicated to study
which of these orientations would in fact be selected by the system.
In this work, we have used a much simpler scheme to demonstrate the
neccessary analysis (albeit within a one-dimensional model- see next
paragraph) to unambiguously determine whether banding actually occurs
in a system, as well as the banding orientation.  While Schmitt
\textit{et al.}  \cite{SchmittMarquesLequeux} have also presented
quite a simple phenomenological model (including both concentration
and momentum as dynamical variables), they did not go beyond a linear
stability analysis. They also did not consider the case (as we have
here) of a non-conserved variable initiating an instability in the
system.

Our calculations have been carried out only for the case of planar
flow.  It has been shown for the Johnson-Segalman model \cite{Olmsted
  JS} that the nonuniformity of stress in a curved geometry has
significant effects on banding. In addition, we have examined the
issue of gradient versus vorticity banding using a one-dimensional
model. Realistically, the model should be solved considering both
vorticity and gradient directions simultaneously.  The band
orientation may be influenced by anisotropy in the diffusion
coefficient.  A convective term of the form $v \cdot{\nabla \phi}$
should also be included in the equation of motion. Such a term does
not appear in a one-dimensional shear flow, but it can qualitatively
affect transients in phase separation in two dimensions. 
Finally, noise has been incorporated into our model through
  the initial conditions. While Gaussian noise is present in the
equation of motion through the diffusion term, in a driven
system there may be other noise terms that should be added.

We stress that our phenomenological theory only aims to describe the
general macroscopic physics of flow-induced phase transitions. Details
concerning the underlying structural transformations can only be
probed by more specific microscopic models.


\begin{thebibliography}{99}
\bibitem{mesh points} In the numerical solution of the integro-
differential equation, we observe a small (on the order of a 
few tenths of a percent) variation with mesh-size of the
steady-state selected stress of the banded states.
\end{thebibliography}
\end{document}